\documentclass[twocolumn,aps]{revtex4}
\usepackage{amsmath}
\usepackage{amssymb}
\usepackage{lipsum}
\usepackage{graphicx}
\usepackage{dcolumn}
\usepackage{bm}
\usepackage{xcolor}
\usepackage{epstopdf}

\begin{document}

\preprint{Phys.Rev.B}

\title{Interaction dominated  transport in 2D conductors: from degenerate to partially-degenerate regime.}

\author{G. M. Gusev,$^1$  A. D. Levin,$^1$  E. B. Olshanetsky$^{2,3}$
 Z. D. Kvon,$^{2,3}$ V. M. Kovalev,$^{2,4,5}$ M. V. Entin$^{2,3}$,  and  N. N. Mikhailov$^{2,3}$}

\affiliation{$^1$Instituto de F\'{\i}sica da Universidade de S\~ao
Paulo, 135960-170, S\~ao Paulo, SP, Brazil}
\affiliation{$^2$Institute of Semiconductor Physics, Novosibirsk
630090, Russia}
\affiliation{$^3$Novosibirsk State University, Novosibirsk 630090,
Russia}
\affiliation{$^4$Novosibirsk State Technical University, Novosibirsk 630073,
Russia}
\affiliation{$^5$Abrikosov Center for Theoretical Physics, Moscow Institute of Physics and Technology, Dolgoprudny, 141701, Russia}

\date{\today}
\begin{abstract}
In this study, we investigate the conductivity of a two-dimensional (2D) system in HgTe quantum well comprising two types of carriers with linear and quadratic spectra, respectively. The interactions between the two-dimensional Dirac holes and the heavy holes lead to the breakdown of Galilean invariance, resulting in interaction-limited resistivity. Our exploration of the transport properties spans from low temperatures, where both subsystems are fully degenerate, to higher temperatures, where the Dirac holes remain degenerate while the heavy holes follow Boltzmann statistics, creating a partially degenerate regime. Through a developed theory, we successfully predict the behavior of resistivity as $\rho\sim T^2$ and $\rho\sim T^{3}$ for the fully degenerate and partially degenerate regimes, respectively, which is in reasonable agreement with experimental observations. Notably, at elevated temperatures, the interaction-limited resistivity surpasses the resistivity caused by impurity scattering by a factor of 5-6. These findings imply that the investigated system serves as a versatile experimental platform for exploring various interaction-limited transport regimes in two component plasma.

\end{abstract}

\maketitle

\section{Introduction}
Recent advances in the fabrication of ultra clean two-dimensional (2D)
semiconductor systems make it possible to explore the regime, where the
interparticle collisions dominate over the impurity  and phonon scattering.
However, in conventional systems with a  parabolic spectrum, where the
Umklapp scattering is prohibited or ineffective due to small Fermi surface,
the particle-particle scattering does not contribute to the conductivity
because it does not change the net current. The situation is changed in
systems with a constrained geometry, where electron transport can be
described by the laws of hydrodynamics. In this case the electron velocity
profile resembles a Poiseuille-like (parabolic) profile and the transport is
governed by electron-electron interactions \cite{gurzhi, polini, narozhny}.
As it was found in the pioneering theoretical study by Gurzhi \cite{gurzhi},
in this regime the particle-particle collisions manifest themselves as a
contribution to resistivity $\sim T^{-2}$, which, however, is essentially
impossible to observe in conventional 2D system  where electron-phonon and
electron-impurity collisions are the dominant scattering mechanisms.
Manifestation of Gurzhi effect can be found in GaAs \cite{dejong, gusev1,
gusev2, gusev3, gusev4} and in graphene \cite{kumar} systems.

\begin{figure*}[ht]
\centering
\includegraphics[width=15cm]{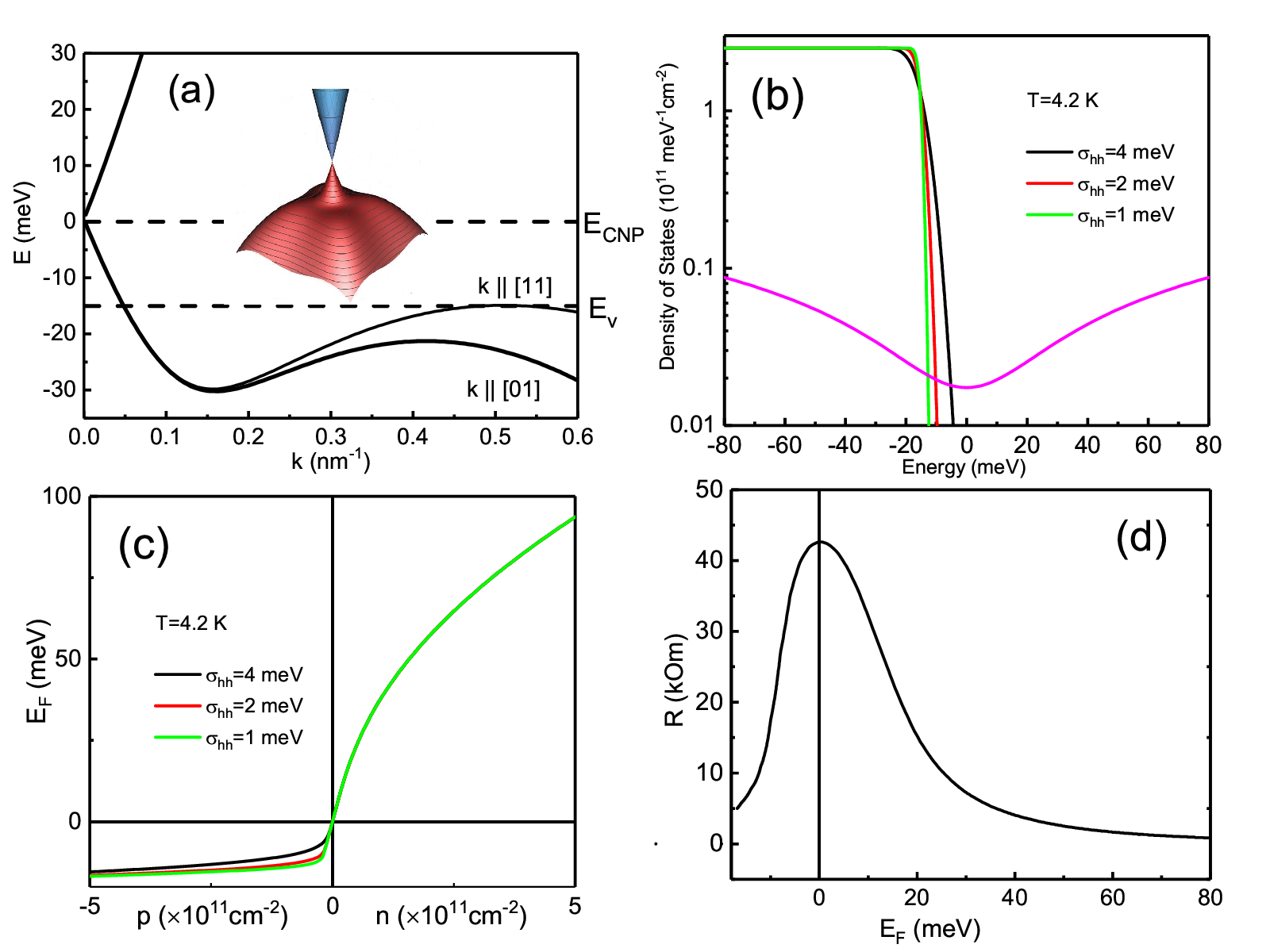}
\caption{(Color online) (a) Schematic representation of the energy spectrum
of a 6.4 nm mercury telluride quantum well. Insert demonstrates schematic 3D
presentation of the spectrum. (b) Density of the states of the Dirac
carriers (magenta) and heavy holes for different values of the heavy hole broadening
parameter $\sigma_{hh}$. (c) The Fermi energy as a function of the charge
carriers density. (d) The resistance of the 6.4 nm sample as a function of
the Fermi energy for $\sigma_{hh}=2 meV$  at T=4 K.}
\end{figure*}

Another important 2D system where particle -particle interaction contributes
to the conductivity is the degenerate and/or nondegenerate electron-hole
plasma in semimetals. In general, a system with a simple parabolic spectrum
is a Galilean invariant electron liquid, while, as expected, a system with
two types of charge carriers that differ in sign and/or effective mass lacks
the Galilean invariance \cite{hwang, nagaev1, nagaev2}. Indeed in
non-Galilean-invariant liquids the colliding carriers  have either opposite
charge sign or different spectra, so the net current is not proportional to
the total momentum of particles. The strong mutual friction between electrons
and holes in a degenerate 2D semimetal leads to the resistivity $\sim T^{2}$,
which has been observed in HgTe quantum wells \cite{olshanetsky,
entin}. The nodegenerate limit has been explored in a single layer and
bilayer graphene, where  e-h pairs are thermally excited \cite{nam, tan}. The
electron-hole collisions rate is expected to be proportional to $T$,
which leads to temperature independent conductivity. Electron-hole friction regime
leading to $T^2$ resistivity has been explored recently in graphene bilayer with spatially separated electrons and holes \cite{bandurin}.

An interesting and important physics is expected to take place when plasma is
partially degenerate: electrons are distributed according to the Fermi statistics,
while the ions obey the Bolztmann statistics \cite{lampe}. The physics associated
with the hydrodynamics in a partially degenerate regime  has not been
explored yet.

In this research, we present the gapless HgTe quantum well as a versatile platform that hosts two subbands with significantly distinct effective masses. HgTe-based devices offer an exceptional opportunity for investigating transitions between various regimes, as they can be precisely adjusted using gate voltage to control density and system degeneracy.

The HgTe quantum well spectrum depends strongly on the well width \cite{gerchikov, kane, bernevig}. At the critical well width $ d_{c}=6.3 - 6.5 $ nm (depending on
the surface orientation and the quantum well deformation), the band gap
becomes zero. The HgTe QWs with the width larger than $ d_{c}$ are the
two-dimensional (2D) topological insulator
\cite{konig, gusev6} and  2D semimetal  \cite{olshanetsky, entin}. .

The energy spectrum of a gapless HgTe quantum well with the width of
$6.3-6.5$ nm consists of a single valley Dirac cone leading to some unusual
electron transport properties \cite{buttner, kozlov, gusev5, kristopenko,
kvon, shuvaev}. At the same time, besides the Dirac-like holes in the center
of the Brillion zone, the valance band has lateral heavy hole valleys
located some distance below the Dirac point (Fig.1a). It may be expected that
the coexistence of the 2D Dirac-like  holes and the heavy holes at a finite
wave vector $k$ will result in the resistance increasing with temperature
since the low mobility heavy holes will change the electron momentum during
collisions. Therefore the gapless HgTe well allows us to study the scattering
between the conventional and the Dirac particles.

In the present paper we report the experimental and theoretical study of the transport in a
gapless HgTe quantum well with the width of $d_{c}=6.3-6.5$ nm. In the region
where the holes with the linear and the parabolic spectrum coexist, the
system lacks the Galilean invariance and we observe resistivity  $\rho \sim
T^2$ in the fully degenerate and $\rho \sim T^\alpha$ ($\alpha \approx 3$) in a partially degenerate
regimes, where the heavy holes obey the Boltzman statistic, while the Dirac
holes remain degenerate. Such T dependence is attributed to the mutual
friction between the Dirac and the heavy holes.

\section{Electron spectrum in a gapless HgT-base quantum well}
To understand the transport properties of the charge carriers in a gapless
HgTe quantum well we first show the energy spectrum in a wide energy interval
for both conductance and the valence bands in figure 1a. Such HgTe well hosts
the Dirac fermions with a linear electron and holes spectrum
$\varepsilon_e=\pm v|k|$, where the Fermi velocity is  $
v=7\times10^{7} cm/s = c/430$ (c is the light velocity), k is momentum.   In addition one can see the lateral maximum of the valence band below the charge neutrality
point: $k_0$ is defined as $vk_0=\mu$, $p_0=m_h v_F=\sqrt{2m_h(\mu-\Delta)}$, $v_F$ is the Fermi velocity of the heavy holes,  $m_h=0.15 m_{0}$ is the effective mass of the holes, $\mu$ is the electrochemical potential and  $\Delta \approx 15 meV$, where  $\Delta=|E_{v}-E_{CNP}|$. The band structure has been computed in various previous studies \cite{buttner, kristopenko}. Specifically, we employed the model outlined in our earlier paper \cite{raichev}, where comprehensive details of the calculations were provided. Figures 1 b and c show  the density of states (DOS) for Dirac
carriers and heavy holes and the Fermi energy at T=4.2 K  respectively for
different heavy hole broadening parameter $\sigma_{hh}$. Because of the large
effective mass and the valley degeneracy $g_{v}=2$  the density of states of
the heavy holes (HH) is much larger than the density of states of the Dirac
holes (DH). In realistic samples  the in-plane fluctuations of the QW width
about its average value $d\approx d_{c}$, that cannot be  avoided during the
QW growth, lead to in-plane variations of the bands gap \cite{gusev8}, and,
consequently, to variations of the charge neutrality point (CNP) and $E_{v}$ positions. This results in
a broadening of the density of states, shown in the fig.1b. From topological
network model, described in ref. \cite{gusev8} we estimate the DOS broadening
$\sim 1-4 meV$. The Fermi energy pinning in the tail of the
heavy-hole DOS ( Fig. 1 c) may lead to a strong asymmetry of the $R(V_{g})$ dependence in contrast  to graphene.  Figure 1d demonstrates the resistance as a function
of the Fermi energy for one of the typical gapless HgTe QW sample. The Fermi
level crosses $E_{v}$ for parameters $\sigma_{hh}=1-4 meV$. For large
broadening $\sigma_{hh}=6 meV$ the Fermi level does not cross $E_{v}$ because
it gets pinned in the disorder induced heavy holes DOS tail ( figure 1 c) and
the Dirac holes move through isolated islands or puddles where HH and DH
coexist. Below for simplicity we consider a homogeneous two component
conductance model.
\begin{figure*}[ht!]
\centering
\includegraphics[width=17cm]{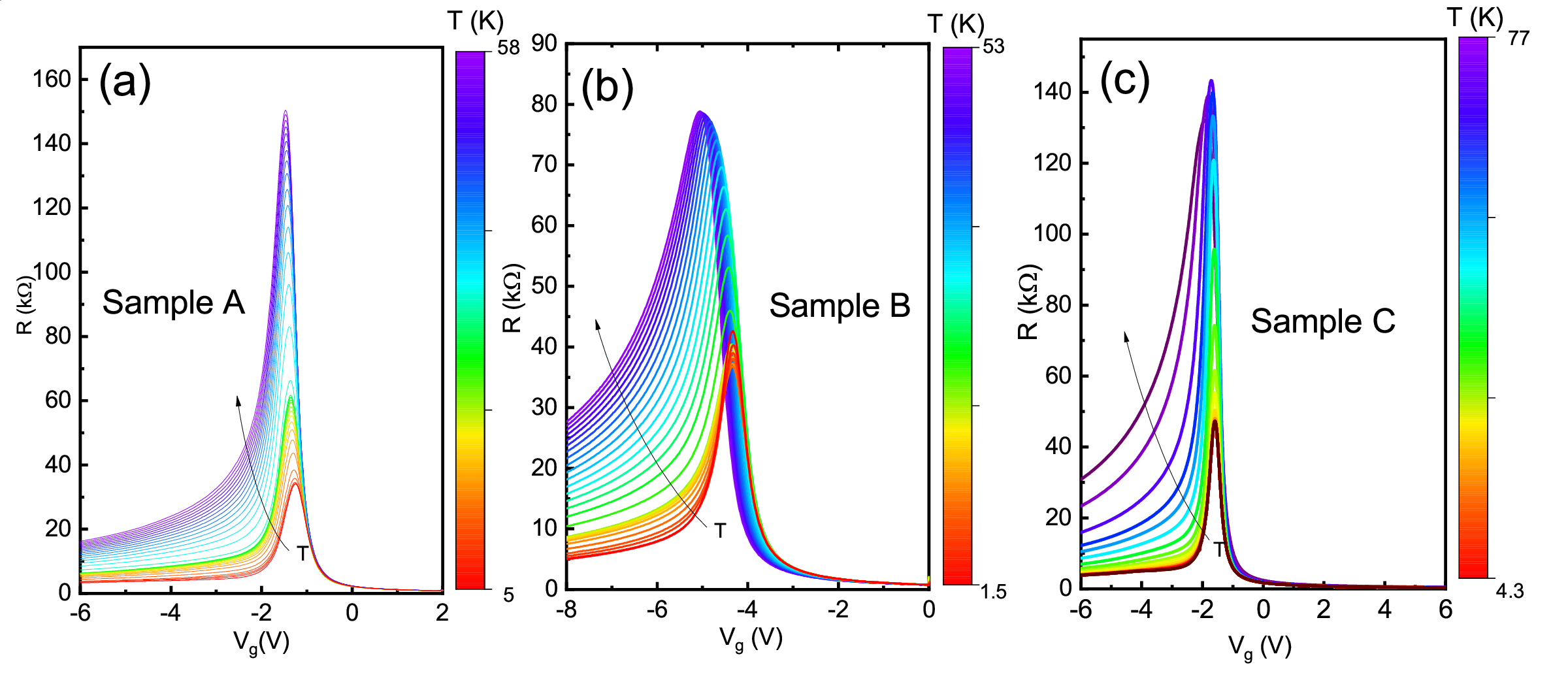}
\caption{(Color online) Resistance as a function of the gate voltage at
different temperatures for three HgTe gapless quantum wells, d=6.4 nm, device
A (a), d=6.3 nm , device B(b), and d=6.4 nm , device C (c). }
\end{figure*}
\section{Sample description and methods}

We have measured the resistance of quantum wells
$Cd_{0.65}Hg_{0.35}Te/HgTe/Cd_{0.65}Hg_{0.35}Te$ with (013) surface
orientations and the well width of 6.3-6.5\:nm. The samples were grown by
means of Molecular Beam Epitaxy at $T \approx 160 - 200^{\circ} C$  on GaAs
substrate with the (013) surface orientation \cite{mikhailov}. Such inclined
to 100 orientation substrates have been used in order to increase the quality
of the material. Layer sequence scheme and details of the sample preparation
has been published previously. The experimental devices were
Hall bars with eight voltage probes divided into 3 separate segments with the
width $W$ of $50\text{\:$\mu$m}$ and the lengths $L$ $(100, 250,
100\text{\:$\mu$m})$ between the probes. A dielectric layer (200\:nm of
$SiO_{2}$) was deposited on the sample surface followed by a TiAu gate
electrode. Ohmic contacts to the quantum well has been fabricated by
annealing of indium on the device contact pads. The sheet density variation
with gate voltage was $1.1\times10^{11}$\text{\:cm$^{-2}$V$^{-1}$}. Because
of the dielectric breakdown the limiting gate voltage was $\pm 8 V$. The
measurements of the resistance $R(T)$  has been performed in the temperature
range 4.2 - 70\:K  using a conventional 4 probe set up with a 1-27\:Hz ac
current of 1-10\:nA through the sample.
\section{Experimental results}

\begin{table}[ht]
\centering
\begin{tabular}{|l|l|l|l||l||l||l|}
\hline
sample & $d$ (nm) & $V_{CNP}$ (V)& $\rho_{max}(h/e^{2})$& $\mu_{e} (V/cm^{2}s )$ \\
\hline
A&    6.4 & -1.3& 0.26 & 96.000\\
\hline
B&   6.3 & -4.3 & 0.33 &63.000\\
\hline
C&   6.4 & -1.6 & 0.37 &105.600 \\
\hline
\end{tabular}
\caption{\label{tab1} Some of the typical parameters of the electron system in HgTe quantum well at T=4.2K.}
\end{table}

Figure 2 presents the resistance as a function of the gate voltage obtained
over a wider temperature range  for three different samples A, B and C. The table
1 lists the typical parameters of the gapless HgT quantum well, used in this
study, such as the well width d, the gate voltage corresponding to the Dirac
point position $V_{CNP}$, the resistivity ($\rho$ value at the CNP  and the
electron mobility $\mu_{e} = 1/\rho n_{s}$ for the electron density $n_{s} = 10^{11}
cm^{-2}$. The evolution of the resistance with temperature is similar in all
samples:   at the hole side of the energy spectrum the resistance increases
with T, indicating a metallic type of conductivity. Moreover, with  T
increasing the resistance peak  shifts to the hole side and becomes wider, while the resistance
at the electronic side remains temperature independent.

In order to analyse the functional form of the resistance (resistivity)
dependence, it could be instructive to subtract T-independent contribution
to  $\rho(T)$ and calculate the excess resistivity $\Delta
\rho(T)=\rho(T)-\rho(T=4.2K)$.  We plot the experimentally measured
excess resistivity $\Delta \rho(T)$ as a function of temperature for all three
devices and at the highest negative gate voltages in  figures 3a,b,c. One can see, that
$\Delta \rho(T)$  exhibits peculiar $T^{2}$ growth, varying by almost two
orders of magnitude. At temperatures $T > 20-30 K$ $\Delta \rho(T)$ reveals
a cubic rather than quadratic temperature dependence. A stronger T dependence at
high temperatures is related to the transition from the fully degenerate to a partially
degenerate regime, where the heavy holes obey the Boltzmann statistic. At the highest
negative gate voltage we achieve the chemical potential
$\mu-\Delta\approx2 meV$ with corresponding Fermi temperature
$T_{F}=(\mu-\Delta)/k\approx 30 K$ (see figure 3a), while the
chemical potential $\mu$ is $\mu\approx17 meV$, $k$ is the Boltzmann constant
\begin{figure*}
  \centering
\includegraphics[width=15cm]{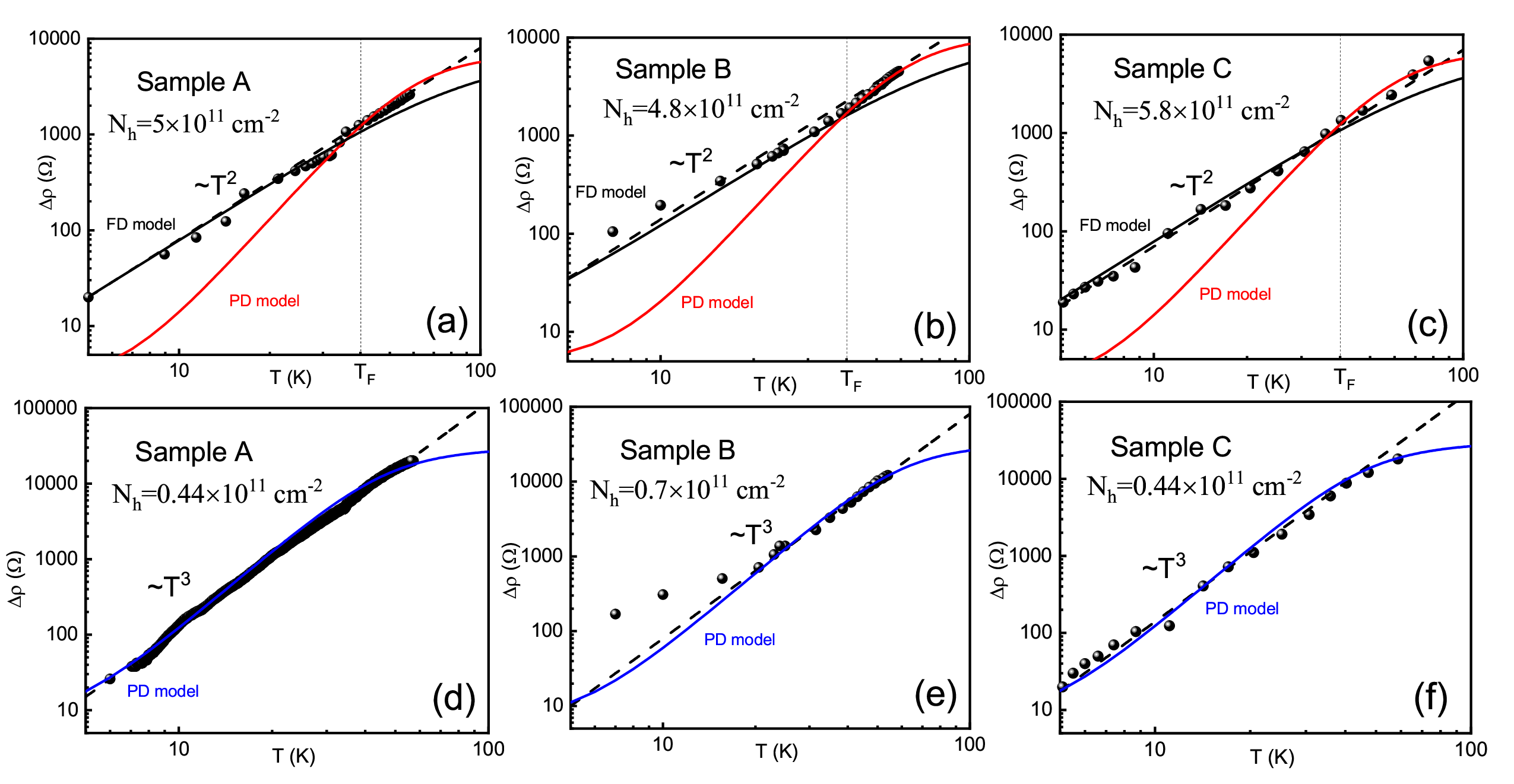}
\caption{(Color online) Excess resistivity $\Delta
\rho(T)=\rho(T)-\rho(T=4.2K)$ as a function of the temperature for high (a,b,c)
and low  (d,e,f) densities for three HgTe quantum wells. (a) The total density is
$N_{h}=5\times10^{11} cm^{-2}$ (sample A); (b) $4.8\times10^{11} cm^{-2}$ (sample
B); (c) $5.8\times10^{11} cm^{-2}$ (sample C). Vertical line indicates
$T_{F}$ Fermi temperature of the heavy holes. (d) The total
density is $N_{h}=0.44\times10^{11} cm^{-2}$ (sample A); (e) $0.7\times10^{11}
cm^{-2}$ (sample B); (f) $0.44\times10^{11} cm^{-2}$ (sample C). Circles represents the experimental data, the blue lines represent the theory of interparticle scattering between  the fully degenerate particles (FD model), the red lines represent the theory of interparticle scattering between fully the degenerate Dirac holes non-degenerate heavy holes (PD model), dashes represents $T^{2}$  (a,b,c) and $T^3$ (d,e,f) dependencies. }
\end{figure*}
Figs 3d,e,f shows the $\Delta \rho(T)$ dependencies at a low total carrier density,
where the partially degenerate regime is expected over the whole temperature
interval. One can see that the experimental data closely follows $\Delta
\rho(T)\sim T^{3}$  dependence.

The peculiar $T^{2}$ and $T^{3}$ resistance dependencies are the
signature of a particle-particle scattering, because the scattering by phonons
would result in a linear rather than quadratic T-dependence \cite{melezhik}.
While it is well-known that the electron-electron  interaction cannot affect
the resistivity of a Galilean-invariant Fermi liquid, there are situations
where a conductivity $\sim T^{-2}$ \cite{pal} may be observed, such as
a multiply connected Fermi surface \cite{pal}, the presence of spin-orbit
interaction \cite{nagaev1}, or several subbands \cite{appel, murzin,
kravchenko, nagaev2, pal}.

In concluding part of this section we would like to discuss the potential impact of temperature on the band structure of HgTe quantum wells \cite{laurenti, becker, li}. Notably, this impact becomes more apparent at elevated temperatures beyond $70$ K. The manifestation of the quantum Hall effect and Shubnikov-de Haas (SdH) oscillations in HgTe wells at $T > 70$ K suggests a limited sensitivity of transport properties to higher temperatures \cite{khouri,kozlov}. Neither the Shubnikov-de Haas oscillations nor the Hall effect undergo significant changes at $T < 70$ K. It is noteworthy that we exclusively observed $T^2$ and $T^3$ dependencies for hole contributions to resistivity, with no corresponding dependencies for electrons. Emphasizing the importance of band gap opening, we expect an activation-type temperature dependence of resistivity, consistent with our observations in samples with an 8 nm well width exhibiting a gap \cite{gusev6}. The diverse experiments carried out by various research groups, coupled with recent theoretical calculations\cite{kristopenko2}, indicate that providing a definitive demonstration of a significant temperature impact on the gap in HgTe wells below $70$ K remains elusive. Furthermore, the presence of the gap leads to an opposite temperature dependence of resistance, wherein resistance typically decreases with increasing temperature. In contrast, our observations in the sample indicate an increase in resistance with rising temperature (fig 2) \cite{gusev6, raichev}.
\section{Two subband hydrodynamic model.}

Below we consider a simple two-subband hydrodynamic model, where the hole-hole scattering is described in terms of a mutual friction. The conductivity can be found
from the equation of motion for the holes with Dirac dispersion and the heavy holes :
\begin {equation}\label{eq1}
-\frac{v_d-v_h} {\tau_{dh}}  - \frac {v_d} {\tau_d} + \frac{q E}{m_d}= 0;
-\frac{v_h-v_d} {\tau_{hd}}  -\frac{v_h}{\tau_h} + \frac{q E}{m_h}=0;
\end {equation}
\begin{widetext}
\begin {equation}\label{eq2}
v_d=\frac{E q \tau _d \left(m_d \tau _h \tau _{\text{hd}}+\tau _{\text{dh}} m_h \left(\tau _h+\tau _{\text{hd}}\right)\right)}{m_d m_h \left(\tau _d \tau _{\text{hd}}+\tau _{\text{dh}} \left(\tau _h+\tau _{\text{hd}}\right)\right)};
v_h=\frac{E q \tau _h \left(\tau _d \tau _{\text{dh}} m_h+m_d \tau _{\text{hd}} \left(\tau _d+\tau _{\text{dh}}\right)\right)}{m_d m_h \left(\tau _d \tau _{\text{hd}}+\tau _{\text{dh}} \left(\tau _h+\tau _{\text{hd}}\right)\right)};
\end {equation}
\end{widetext}
where  E is the electric  field, $v_{h,d}$ are the carrier velocities for heavy (h) and Dirac (d) holes, $m_{h}$ is heavy hole effective mass, $m_{d}=\mu/v^{2}$ is Dirac hole effective mass, $\tau_{h(d)}$ are the momentum relaxation times,  $\tau_{h d (d h)}$ is collision time with heavy (Dirac) holes per Dirac (heavy) holes.
\begin {equation}\label{eq3}
\tau_{dh}=\frac{\tau _{\text{int}} \left(m_d n_d+m_h n_h\right)}{m_h n_h};
\tau _{hd}=\frac{\tau _{\text{int}} \left(m_d n_d+m_h n_h\right)}{m_d n_d};
\end {equation}
One can introduce  the center-of-mass  velocity and the relative velocity:
\begin {equation}\label{eq4}
U=\frac{m_d n_d v_d+m_h n_h v_h}{m_d n_d+m_h n_h};
V=v_d-v_h.
\end {equation}
Resolving equations \ref{eq1} - \ref{eq4} we obtain:
\begin{widetext}
\begin {equation}\label{eq5}
U=\frac{q E \left(n_h \tau _h \left(\tau _d+\tau _{\text{int}}\right)+n_d \tau _d \left(\tau _h+\tau _{\text{int}}\right)\right)}{m_h n_h \left(\tau _d+\tau _{\text{int}}\right)+m_d n_d \left(\tau _h+\tau _{\text{int}}\right)};
\end {equation}

\begin {equation}\label{eq6}
V=\frac{q E \left(m_d n_d + m_h n_h\right) \left(\tau_d m_h-m_d \tau_h \right)}
{m_d m_h \left(m_h n_h \left(\tau _d+\tau _{int} \right)+m_d n_d \left(\tau _h+\tau _{\text{int}}\right)\right)}\tau_{int} .
\end {equation}
\end{widetext}
where $1/\tau_{int}=1/\tau_{d h}+1/\tau_{h d}$.  This equation considering the constraint, that interactions between holes conserve the overall momentum density:
$n_d m_d \tau_{d h}=n_h m_h \tau_{h d}$.
The initial term signifies the impact arising from the movement of the Dirac and heavy holes in opposing directions due to the contrasting forces imposed on them by the electric field. On the other hand, the second term indicates the influence resulting from Dirac  and heavy holes collectively moving in the same direction due to the Coulomb drag, analogous to a "friction," occurring between them.
\begin {equation}\label{eq7}
j=\left(n_d+n_h\right) q U + \frac{ n_d n_h}{m_d n_d+m_h n_h}\left(m_d+m_h\right)q V
\end {equation}
\begin{widetext}
\begin {equation}\label{eq8}
j=\left[\frac{ q^2 \left(n_d+n_h\right) \left(n_d \tau _d \left(\tau _h+\tau _{int}\right)+n_h \tau _h \left(\tau _d+\tau _{int}\right) \right)}
{m_h n_h \left(\tau _d+\tau _{\text{int}}\right)+m_d n_d \left(\tau _h+\tau _{\text{int}}\right)}\right]E
-\left[\frac{ q^2 n_d n_h \tau _{\text{int}} \left(m_d+m_h\right) \left(m_d \tau _h-\tau _d m_h\right)}{m_d m_h \left(m_h n_h \left(\tau _d+\tau _{\text{int}}\right)+m_d n_d \left(\tau _h+\tau _{\text{int}}\right)\right)}\right]E
\end{equation}
\begin {equation}\label{eq9}
\sigma=q^2 \frac{n_d n_h\left[  \tau_d  \left(\tau_h+\tau _{int}\right)\left(2+ \frac{n_d}{n_h}\right)+
\tau _h \left(\tau_d+\tau _{int}\right)\frac{n_h}{n_d}+\tau _{int}\left(\frac{m_h}{m_d} \tau_d-\frac{m_d}{m_h}\tau _h\right)\right]}
{m_h n_h \left(\tau _d+\tau _{int}\right)+m_d n_d \left(\tau _h+\tau _{\text{int}}\right)}
\end{equation}
\end{widetext}
Where we used the expression for the conductivity  $j=\sigma E$.

One might anticipate that the scattering times $\tau_{d(h)}$ associated with impurity and interface roughness scattering remain temperature-independent. On the other hand, the scattering time $\tau_{int}$ attributed to hole-hole friction is accountable for the temperature-dependent contribution to conductivity in the non-Galilean hole liquid.

It would be instructive to consider $T=0$ and $T=\infty$ limits in the case when the effective masses are significantly different. The Dirac hole effective scattering time can be estimated from relation $m_{d}=\mu/v^{2}\approx0.006 m_{0}$, at $\mu\approx 16 meV$.

At $T\rightarrow 0$ both bands contribute to the total conductivity dominated by Dirac holes:
\begin{equation}\label{eq10}
\sigma(T=0)=\frac{q^{2}n_{d}v^{2}\tau_{d}}{\mu}=\frac{q^{2}n_{d}\tau_{d}}{m_{d}}.
\end{equation}

At $T=\infty$ the conductivity becomes temperature independent and saturates at a value approximately determined by the conductivity of the heavy hole band :
\begin{equation}\label{eq11}
\sigma(T=\infty)=\frac{q^{2}(n_{d}+n_{h})^{2}\tau_{h}}{m_{h}n_{h}}
\end{equation}
The ratio of the resistivities at both temperature limits is given by
\begin{equation}\label{eq12}
\frac{\rho(T=\infty)}{\rho(T=0)}=\frac{m _{h}n_{h}n_{d}\tau_{d}}{m_{d}(n_{d}+n_{h})^{2}\tau_{h}}
\end{equation}
In our case $n_{h}>>n_{d}$, the ratio of the resistivities at both temperature limits is determined by :
\begin{equation}\label{eq13}
\frac{\rho(T=\infty)}{\rho(T=0)}=\frac{m _{h}n_{d}\tau_{d}}{m_{d}n_{h}\tau_{h}}
\end{equation}

It is important to consider the resistivity at the two opposite temperature limits. If
$n_{d}\tau_{d}\sim n_{h}\tau_{h}$ the ratio of the resistivities corresponding to the two temperature limits is determined by
the effective mass ratio:

$\frac{\rho(T=\infty)}{\rho(T=0)}\approx\frac{m _{h}}{m_{d}}\approx50$.

Thus, impurity scattering is predominant at lower temperatures, while the
hole-hole interaction dominates at higher temperatures.
Here, one can highlight, as evident from Figure 2, that in actual samples, the hydrodynamic conductivity can surpass the Drude conductivity by a factor of 5 to 6. This observation underscores that our system stands out as a liquid with the highest hydrodynamic conductivity when compared to other systems, as discussed in references \cite{polini} and \cite{narozhny}. Consequently, it serves as a promising platform for investigating various hydrodynamic phenomena, such as the violation of the Wiedemann-Franz law, collective sound modes, and nonlinear behaviors (refer to \cite{polini} and \cite{narozhny} for a comprehensive review).

In the next section we consider theoretically mechanism of the scattering between Dirac and heavy holes.

\section{Calculation of the scattering time. Degenerate Dirac and parabolic (heavy) holes}

The formula $\epsilon_{\bf k}=vk$ represents the energy of  Dirac particles, ${\bf k}$ is  the momentum ${\bf k}$, $v$ is velocity. The formula for  particles with parabolic spectrum  is $\epsilon_{\bf p}=p^2/2m+\Delta$, where ${\bf p}$ is the heavy hole momentum, $m=m_h$ is the heavy hole mass.  Let's set the Planck and Boltzmann constants to be 1 for the calculations, and in the final equations, we will revert to its original value.
The kinetic equation for   Dirac particles is given by:
\begin{widetext}
\begin{gather}\label{eq14}
({\bf F},{\bf v}_{\bf k})\frac{\partial n_{\bf k}}{\partial \epsilon_{\bf k}}=2\pi\sum_{{\bf p}',{\bf k}',{\bf p}}|U_{{\bf p}'-{\bf p}}|^2
[(1-f_{\bf k})(1-f_{\bf p})f_{{\bf k}'}f_{{\bf p}'}-
(1-f_{{\bf k}'})(1-f_{{\bf p}'})f_{{\bf k}}f_{{\bf p}}]
\delta(\epsilon_{{\bf k}'}+\epsilon_{{\bf p}'}-\epsilon_{{\bf k}}-\epsilon_{{\bf p}})
\delta_{{\bf k}'+{\bf p}'-{\bf k}-{\bf p}}.
\end{gather}
\end{widetext}

For heavy hole particles with parabolic spectrum the kinetic equations is:
\begin{widetext}
\begin{gather}\label{eq15}
({\bf F},{\bf v}_{\bf p})\frac{\partial n_{\bf p}}{\partial \epsilon_{\bf p}}=2\pi\sum_{{\bf p}',{\bf k}',{\bf k}}|U_{{\bf p}'-{\bf p}}|^2
[(1-f_{\bf k})(1-f_{\bf p})f_{{\bf k}'}f_{{\bf p}'}-
(1-f_{{\bf k}'})(1-f_{{\bf p}'})f_{{\bf k}}f_{{\bf p}}]
\delta(\epsilon_{{\bf k}'}+\epsilon_{{\bf p}'}-\epsilon_{{\bf k}}-\epsilon_{{\bf p}})
\delta_{{\bf k}'+{\bf p}'-{\bf k}-{\bf p}}.
\end{gather}
\end{widetext}

Here ${\bf v}_{\bf k}=v{\bf k}/k$ and ${\bf v}_{\bf p}={\bf p}/m$  are velocity of the Dirac and parabolic holes, consequently. We will linearize these equations with small corrections to the distribution functions: $f_{\bf k}=n_{\bf k}+\delta f_{\bf k}$ and $f_{\bf p}=n_{\bf p}+\delta f_{\bf p}$, where $n_{\bf k}$ and $n_{\bf p}$ represents equilibrium distributions of Dirac and parabolic holes. Assuming that in the kinetic equation for Dirac holes, the parabolic holes are in equilibrium (and vice versa), we obtain:

\begin{widetext}
\begin{gather}\label{eq16}
({\bf F},{\bf v}_{\bf k})\frac{\partial n_{\bf k}}{\partial \epsilon_{\bf k}}=
-2\pi\sum_{{\bf p}',{\bf k}',{\bf p}}|U_{{\bf p}'-{\bf p}}|^2
\Bigl[\delta f_{\bf k}[(1-n_{\bf p})n_{{\bf k}'}n_{{\bf p}'}+n_{\bf p}(1-n_{{\bf k}'})(1-n_{{\bf p}'})]-\\\nonumber
-\delta f_{{\bf k}'}[(1-n_{{\bf k}})(1-n_{{\bf p}})n_{{\bf p}'}+n_{\bf k}n_{\bf p}(1-n_{{\bf p}'})]\Bigr]
\delta(\epsilon_{{\bf k}'}+\epsilon_{{\bf p}'}-\epsilon_{{\bf k}}-\epsilon_{{\bf p}})
\delta_{{\bf k}'+{\bf p}'-{\bf k}-{\bf p}},
\end{gather}
\begin{gather}\label{eq17}
({\bf F},{\bf v}_{\bf p})\frac{\partial n_{\bf p}}{\partial \epsilon_{\bf p}}=
-2\pi\sum_{{\bf p}',{\bf k}',{\bf k}}|U_{{\bf p}'-{\bf p}}|^2
\Bigl[\delta f_{\bf p}[(1-n_{\bf k})n_{{\bf k}'}n_{{\bf p}'}+n_{\bf k}(1-n_{{\bf k}'})(1-n_{{\bf p}'})]-\\\nonumber
-\delta f_{{\bf p}'}[(1-n_{{\bf k}})(1-n_{{\bf p}})n_{{\bf k}'}+n_{\bf k}n_{\bf p}(1-n_{{\bf k}'})]\Bigr]
\delta(\epsilon_{{\bf k}'}+\epsilon_{{\bf p}'}-\epsilon_{{\bf k}}-\epsilon_{{\bf p}})
\delta_{{\bf k}'+{\bf p}'-{\bf k}-{\bf p}}.
\end{gather}
\end{widetext}

The corrections to the distribution functions can be presented in the form :

\begin{gather}\label{eq18}
\delta f_{\bf k}=\frac{\partial n_{\bf k}}{\partial \epsilon_{\bf k}}\varphi_{\bf k}=-\frac{\varphi_{\bf k}}{T}n_{\bf k}(1-n_{\bf k}),\\\nonumber
\delta f_{\bf p}=\frac{\partial n_{\bf p}}{\partial \epsilon_{\bf p}}\varphi_{\bf p}=-\frac{\varphi_{\bf p}}{T}n_{\bf p}(1-n_{\bf p}).
\end{gather}

However, it is important to take into account the following auxiliary relations:

\begin{widetext}
\begin{gather}\label{eq19}
\delta(\epsilon_{{\bf k}'}+\epsilon_{{\bf p}'}-\epsilon_{{\bf k}}-\epsilon_{{\bf p}})=
\int d\omega\delta(\epsilon_{{\bf k}'}-\epsilon_{{\bf k}}-\omega)\delta(\epsilon_{{\bf p}'}-\epsilon_{{\bf p}}+\omega),\\\nonumber
\delta_{{\bf k}'+{\bf p}'-{\bf k}-{\bf p}}=\sum_{\bf q}\delta_{{\bf k}'-{\bf k}-{\bf q}}\delta_{{\bf p}'-{\bf p}+{\bf q}},\\\nonumber
(1-n_{\bf p})n_{{\bf p}'}=N_{-\omega}(n_{\bf p}-n_{{\bf p}'})=-(1+N_\omega)(n_{\bf p}-n_{{\bf p}'}),\\\nonumber
(1-n_{\bf k})n_{{\bf k}'}=N_{\omega}(n_{\bf k}-n_{{\bf k}'}),\\\nonumber
N_\omega=\frac{1}{e^{\omega/T}-1},\,\,\,\,
\frac{\partial N}{\partial \omega}=\frac{N_{-\omega}N_{\omega}}{T}=-\frac{N_{\omega}(1+N_\omega)}{T}.
\end{gather}
\end{widetext}

After some algebra, equations \eqref{eq18} and \eqref{eq19} can be presented in form

\begin{widetext}
\begin{gather}\label{eq20}
({\bf F},{\bf v}_{\bf k})\frac{\partial n_{\bf k}}{\partial \epsilon_{\bf k}}=
2\pi\sum_{{\bf p}',{\bf k}',{\bf p},{\bf q}}\int d\omega|U_{{\bf p}'-{\bf p}}|^2
(\varphi_{\bf k}-\varphi_{{\bf k}'})\frac{\partial N}{\partial\omega}(n_{\bf p}-n_{{\bf p}'})(n_{\bf k}-n_{{\bf k}'})\times\\\nonumber\times
\delta(\epsilon_{{\bf k}'}-\epsilon_{{\bf k}}-\omega)\delta(\epsilon_{{\bf p}'}-\epsilon_{{\bf p}}+\omega)
\delta_{{\bf k}'-{\bf k}-{\bf q}}\delta_{{\bf p}'-{\bf p}+{\bf q}},\\\nonumber
({\bf F},{\bf v}_{\bf p})\frac{\partial n_{\bf p}}{\partial \epsilon_{\bf p}}=
2\pi\sum_{{\bf p}',{\bf k}',{\bf k},{\bf q}}\int d\omega|U_{{\bf p}'-{\bf p}}|^2
(\varphi_{\bf p}-\varphi_{{\bf p}'})\frac{\partial N}{\partial\omega}(n_{\bf p}-n_{{\bf p}'})(n_{\bf k}-n_{{\bf k}'})\times\\\nonumber\times
\delta(\epsilon_{{\bf k}'}-\epsilon_{{\bf k}}-\omega)\delta(\epsilon_{{\bf p}'}-\epsilon_{{\bf p}}+\omega)
\delta_{{\bf k}'-{\bf k}-{\bf q}}\delta_{{\bf p}'-{\bf p}+{\bf q}}.
\end{gather}
\end{widetext}
The energy transferred between colliding holes, $\omega$, is of the order of temperature, $\omega\sim T$. Thus, in degenerate case the difference of distribution functions can be expanded as:
\begin{gather}\label{eq21}
n_{\bf k}-n_{{\bf k}'}=n(\epsilon_{{\bf k}})-n(\epsilon_{{\bf k}}+\omega)\approx -\omega\frac{\partial n_{\bf k}}{\partial \epsilon_{\bf k}},\\\nonumber
n_{\bf p}-n_{{\bf p}'}=n(\epsilon_{{\bf p}})-n(\epsilon_{{\bf p}}-\omega)\approx \omega\frac{\partial n_{\bf p}}{\partial \epsilon_{\bf p}}.
\end{gather}
Referring to equation \eqref{eq21}, the derivative $\frac{\partial n_{\bf k}}{\partial \epsilon_{\bf k}}$ represents the rate of change of the occupation number $n_{\bf k}$ with respect to the energy $\epsilon_{\bf k}$. It captures the effects and contributions of various scattering processes, including both elastic and inelastic scattering, determining the thermal and transport properties, respectively. Substituting Eq.\eqref{eq20} into
Eqs.\eqref{eq21}, one finds
\begin{widetext}
\begin{gather}\label{eq22}
({\bf F},{\bf v}_{\bf k})=
-2\pi\sum_{{\bf p},{\bf q}}|U_{{\bf q}}|^2(\varphi_{\bf k}-\varphi_{{\bf k}+{\bf q}})
\frac{\partial n_{\bf p}}{\partial \epsilon_{\bf p}}
\int \omega^2d\omega
\frac{\partial N}{\partial\omega}
\delta(\epsilon_{{\bf k}+{\bf q}}-\epsilon_{{\bf k}}-\omega)\delta(\epsilon_{{\bf p}-{\bf q}}-\epsilon_{{\bf p}}+\omega),\\\nonumber
({\bf F},{\bf v}_{\bf p})=
-2\pi\sum_{{\bf k},{\bf q}}|U_{{\bf q}}|^2(\varphi_{\bf p}-\varphi_{{\bf p}-{\bf q}})
\frac{\partial n_{\bf k}}{\partial \epsilon_{\bf k}}
\int \omega^2d\omega
\frac{\partial N}{\partial\omega}
\delta(\epsilon_{{\bf k}+{\bf q}}-\epsilon_{{\bf k}}-\omega)\delta(\epsilon_{{\bf p}-{\bf q}}-\epsilon_{{\bf p}}+\omega).
\end{gather}
\end{widetext}
Due to the degeneracy of both types of holes, one can determine $|{\bf k}|$ from the dispersion relation (assuming isotropic $\partial n_{\bf k}/\partial \epsilon_{\bf k}$): $|{\bf k}|=k_0$, where $k_0$ corresponds to the Fermi surface defined by $vk_0=\mu$. Similarly, we can determine $|{\bf p}|=p_0$ from $p_0^2/2m+\Delta=\mu$ by considering the partial derivative $\partial n_{\bf p}/\partial \epsilon_{\bf p}$.

Let's also introduce the distribution functions corrections $\varphi_{\bf k}$ in the form $\varphi_{\bf k}=({\bf F},{\bf v}_{\bf k})\chi(\epsilon_{\bf k})$, where $\chi(\epsilon_{\bf k})$ is an arbitrary function of energy. For degenerate hole gases, the function $\chi(\epsilon_{\bf k})$ can be considered as a constant: $\chi(\epsilon_{\bf k})\equiv\chi_0\equiv\tau$. Consequently, we have the following expressions:
\begin{gather}\label{eq23}
({\bf F},{\bf v}_{\bf k})=
2\pi\tau_{dh}\sum_{{\bf p},{\bf q}}|U_{{\bf q}}|^2\Bigl({\bf F},({\bf v}_{{\bf k}}-{\bf v}_{{\bf k}+{\bf q}})\Bigr)
\frac{\partial n_{\bf p}}{\partial \epsilon_{\bf p}}
\int \omega^2d\omega
\frac{\partial N}{\partial\omega}
\delta(\epsilon_{{\bf k}+{\bf q}}-\epsilon_{{\bf k}}-\omega)\delta(\epsilon_{{\bf p}-{\bf q}}-\epsilon_{{\bf p}}+\omega),\\\nonumber
({\bf F},{\bf v}_{\bf p})=
2\pi\tau_{hd}\sum_{{\bf k},{\bf q}}|U_{{\bf q}}|^2\Bigl({\bf F},({\bf v}_{{\bf p}}-{\bf v}_{{\bf p}-{\bf q}})\Bigr)
\frac{\partial n_{\bf k}}{\partial \epsilon_{\bf k}}
\int \omega^2d\omega
\frac{\partial N}{\partial\omega}
\delta(\epsilon_{{\bf k}+{\bf q}}-\epsilon_{{\bf k}}-\omega)\delta(\epsilon_{{\bf p}-{\bf q}}-\epsilon_{{\bf p}}+\omega),
\end{gather}
where $\tau_{dh}$ is an Dirac holes scattering time off the heavy holes, whereas $\tau_{dh}$ describes the scattering of heavy holes off the Dirac ones. In the limit of small wave vectors ${\bf q}$, the expressions interring into above expressions can approximated as follows
\begin{widetext}
\begin{gather}\label{eq24}
{\bf F}\cdot({\bf v}_{{\bf k}}-{\bf v}_{{\bf k}+{\bf q}})=v{\bf F}\cdot\left(\frac{{\bf k}}{k}-\frac{{\bf k}+{\bf q}}{|{\bf k}+{\bf q}|}\right)
\approx v\frac{(\textbf{F},\textbf{k})(\textbf{q},\textbf{k})-(\textbf{F},\textbf{q})k^2}{k^3}+
v\frac{(\textbf{F},\textbf{k})[q^2k^2-3(\textbf{q},\textbf{k})^2]+2k^2(\textbf{F},\textbf{q})(\textbf{q},\textbf{k})}{2k^5},
\end{gather}
\end{widetext}

for the Dirac holes, and

\begin{gather}\label{eq25}
{\bf F}\cdot({\bf v}_{{\bf p}}-{\bf v}_{{\bf p}-{\bf q}})=\frac{({\bf F},{\bf q})}{m},
\end{gather}
for heavy ones. Expanding for small momentum transfer also in Delta-functions, we get

\begin{widetext}
\begin{gather}\label{eq26}
v\frac{({\bf F},{\bf k})}{k}=
2\pi\tau_{dh}\sum_{{\bf p},{\bf q}}|U_{{\bf q}}|^2v\left[\frac{(\textbf{F},\textbf{k})(\textbf{q},\textbf{k})-(\textbf{F},\textbf{q})k^2}{k^3}+
\frac{(\textbf{F},\textbf{k})[q^2k^2-3(\textbf{q},\textbf{k})^2]+2k^2(\textbf{F},\textbf{q})(\textbf{q},\textbf{k})}{2k^5}\right]\times\\\nonumber
\times\frac{\partial n_{\bf p}}{\partial \epsilon_{\bf p}}
\int \omega^2d\omega
\frac{\partial N}{\partial\omega}
\delta\left(v\frac{{\bf q}{\bf k}}{k}-\omega\right)\delta\left(-\frac{{\bf p}{\bf q}}{m}+\omega\right),\\\nonumber
\frac{({\bf F},{\bf p})}{m}=2\pi\tau_{hd}\sum_{{\bf k},{\bf q}}|U_{{\bf q}}|^2\frac{({\bf F},{\bf q})}{m}
\frac{\partial n_{\bf k}}{\partial \epsilon_{\bf k}}
\int \omega^2d\omega
\frac{\partial N}{\partial\omega}\delta\left(v\frac{{\bf q}{\bf k}}{k}-\omega\right)\delta\left(-\frac{{\bf p}{\bf q}}{m}+\omega+q^2/2m\right)
\end{gather}
\end{widetext}

Integrating over the angles, we get:
\begin{widetext}
\begin{gather}\label{eq27}
\frac{1}{\tau_{dh}}=
\frac{4\pi }{(2\pi)^4(k_0v)^2}\int qdq|U_{{\bf q}}|^2
\int \omega^2d\omega
\frac{\partial N}{\partial\omega}\int pdp\frac{\partial n_{\bf p}}{\partial \epsilon_{\bf p}}
\frac{\sqrt{(vq)^2-\omega^2}}{\sqrt{(v_pq)^2-\omega^2}},\\\nonumber
\frac{1}{\tau_{hd}}=
\frac{4\pi }{(2\pi)^4 p^2}\int qdq|U_{{\bf q}}|^2
\int \omega^2d\omega
\frac{\partial N}{\partial\omega}
\frac{q^2}{\sqrt{(v_pq)^2-(\omega+q^2/2m)^2}\sqrt{(vq)^2-\omega^2}}\int kdk\frac{\partial n_{\bf k}}{\partial \epsilon_{\bf k}},
\end{gather}
\end{widetext}
where $v_p=p/m_h$. These expressions hold for any type of hole statistics. Below we consider two cases: when Dirac holes are always degenerate, whereas the heavy one are either degenerate or non-degenerate.

\subsection{Degenerate Dirac and Heavy holes}

At low temperatures and hight hole densities, when a Fermi energy is located above $\Delta$, one has $\partial n_{\bf p}/\partial \epsilon_{\bf p}=-\delta(\Delta+\epsilon_{\bf p}-\mu)$
and $\partial n_{\bf k}/\partial \epsilon_{\bf k}=-\delta(vk-\mu)$. Thus, the relation $v_{p=p_0}=v_F=p_0/m$ represents the Fermi velocity of heavy holes, which is a characteristic velocity for fermionic particles. Finally, ignoring the $\omega$ dependence in square roots in Eqs.\eqref{eq27}, and taking into account that
$\int pdp\frac{\partial n_{\bf p}}{\partial \epsilon_{\bf p}}=-m$,
$\int kdk\frac{\partial n_{\bf k}}{\partial \epsilon_{\bf k}}=-\mu/v^2$, one finds
\begin{gather}\label{eq28}
\frac{1}{\tau_{dh}}=-
\frac{4\pi m}{(2\pi)^4(vk_{0})(v_Fk_0)}\left(\int qdq|U_{{\bf q}}|^2\right)
\left(\int \omega^2d\omega
\frac{\partial N}{\partial\omega}\right),\\\nonumber
\frac{1}{\tau_{hd}}=-
\frac{4\pi \mu/v^2}{(2\pi)^4(vp_{0})(v_Fp_0)}\left(\int qdq|U_{{\bf q}}|^2\right)
\left(\int \omega^2d\omega
\frac{\partial N}{\partial\omega}\right)
\end{gather}
\textit{A model of contact interhole interaction.---} Here, we analyze the contact interaction model assuming that $U_{{\bf q}}=U_0$. Integrating as
\begin{widetext}
\begin{gather}\label{eq29}
\int qdq|U_{{\bf q}}|^2=|U_{0}|^2\int\limits_0^{2\min(k_0,p_0)} qdq=2|U_{0}|^2(\min(k_0,p_0))^2,\,\,\,
\int\limits_{-\infty}^{\infty} \omega^2d\omega\frac{\partial N}{\partial\omega}=-\frac{2\pi^2T^2}{3},
\end{gather}
\end{widetext}
we find (recovering Plank and Boltzmann constants)
\begin{gather}\label{eq30}
\frac{1}{\tau_{dh}}=A\frac{m_h (kT)^2|U_0|^2}{3\pi\hbar^5vv_F}\sim T^{2},\\\nonumber
\frac{1}{\tau_{hd}}=\frac{(\mu/v^2) (kT)^2|U_0|^2}{3\pi\hbar^5vv_F}\frac{k_0^2}{p_0^2}\sim T^{2}.
\end{gather}
These formulas are applicable in the framework of QFT. Here: $k_0$ is defined as $vk_0=\mu$, $p_0=m_h v_F=\sqrt{2m_h(\mu-\Delta)}$ for degenerate Dirac and heavy holes gases. It should be noted, that the expressions Eq.\eqref{eq30} satisfy the relation $m_dn_d/\tau_{dh}=m_hn_h/\tau_{hd}$, where $m_d=\mu/v^2$ and $4\pi n_d=k_0^2$ and $4\pi n_h=p_0^2$. The latter expressions hold for degenerate electron gas. A represents a numerical coefficient that varies based on the specifics of the Coulomb interactions.

The contact interaction parameter $U_0$ can be estimated as overscreened Coulomb interaction, $U_0=2\pi e^2/q_s$, where $q_s$ is a screening wave vector.
It is worth noting that this prescription applies to strongly degenerate Fermi liquids. Assuming that the main contribution to the screening is determined by the heavy holes due to its high density of states, the screening wave vector can be written as $q_s=m_he^2/(\epsilon\hbar^2)$, where $\epsilon$ is the dielectric constant of the material.
\begin{figure}[ht!]
\includegraphics[width=10cm]{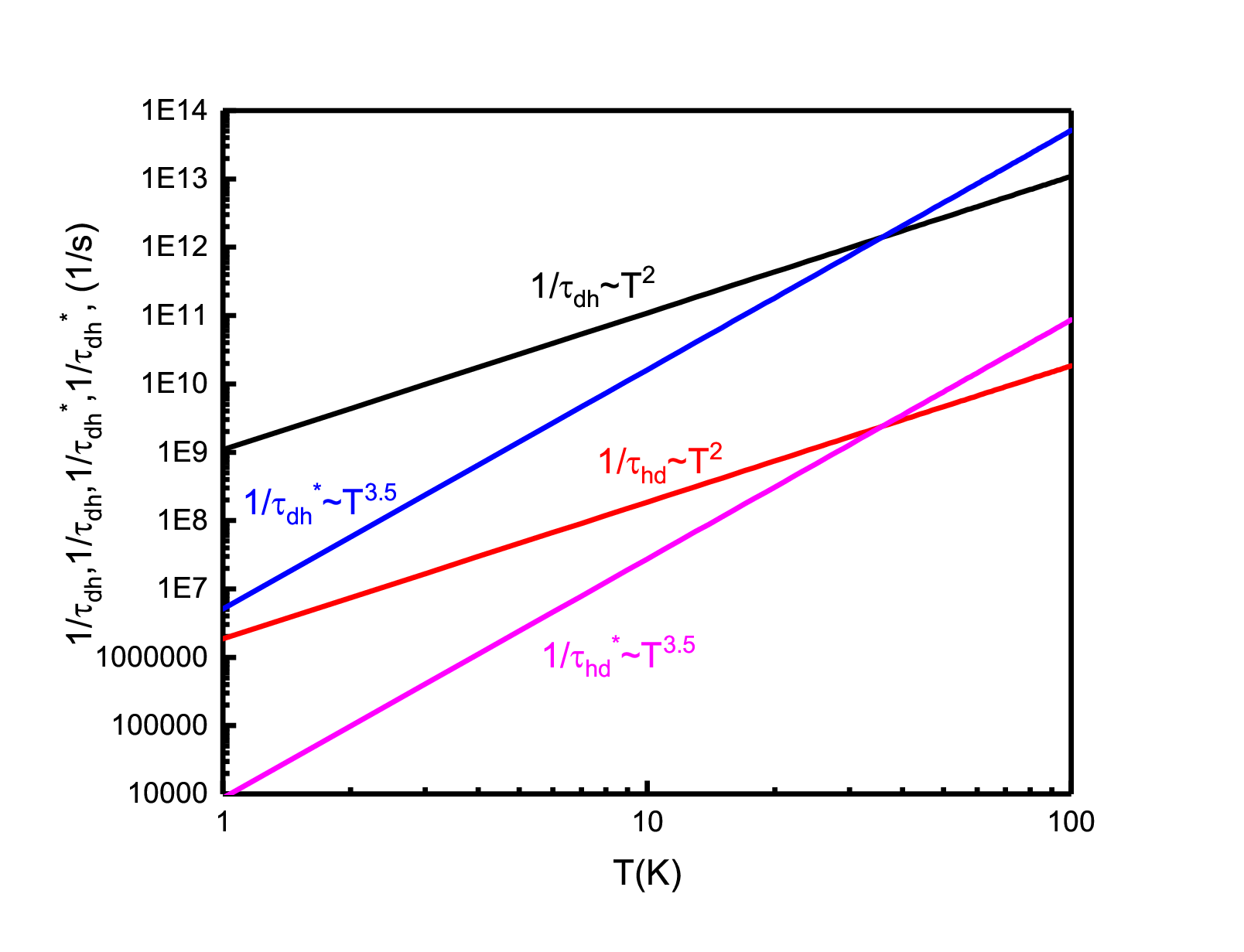}
\caption{(Color online) Temperature dependence of the relaxation rate. The  lines are  computed using equations \ref{eq30} (black), \ref{eq30} (red), \ref{eq32} (blue), \ref{eq32} (magenta). }
\end{figure}

Figure 4 illustrates the relationship between the  relaxation rate $1/\tau_{dh}$ and $1/\tau_{hd}$ and temperature. The values are computed using equations \ref{eq30} with parameter A=1. It is evident that this relationship adheres to a quadratic function of temperature $T^2$. One can see  that $1/\tau_{dh} > 1/\tau_{hd}$.

\subsection{Degenerate Dirac and nondegenerate Heavy holes}

At low densities and moderately hight temperatures, the heavy holes become non-degenerate. In this case the distribution function of heavy holes reads

\begin{gather}\label{eq31}
n_{\bf p}=\frac{2\pi \hbar^2 n_h}{m_h T}\exp\left(-\frac{\epsilon_{\bf p}}{kT}\right).
\end{gather}
Calculations similar to the presented above, yield
\begin{gather}\label{eq32}
\frac{1}{\tau_{dh}^{*}}=B\frac{1.7(kT)^2|U_0|^2n_hm_h}{\hbar^3k_0^2vv_T}\sim T^{7/2},\\\nonumber
\frac{1}{\tau_{hd}^{*}}=\frac{(kT)^2|U_0|^2n_dm_d}{\hbar^3k_0^2vv_T}\sim T^{7/2}.
\end{gather}
 The density of Dirac holes is $4\pi n_d=k_0^2=4\pi \mu^2/v^2=m_d\mu$, $ v_T=\sqrt{2 kT / m}$. The expressions for relaxation times, Eq.\eqref{eq32} also satisfy the general relation $m_dn_d/\tau_{dh}=m_hn_h/\tau_{hd}$. B is a numerical coefficient as a parameter A, contingent upon the nature of Coulomb interactions.

The contact interaction parameter $U_0$ here can also be estimated as overscreened Coulomb interaction, $U_0=2\pi e^2/q_T$, where $q_T$ is a screening wave vector.
Assuming again that the main contribution to the screening is determined by the non-degenerate heavy holes due to its high density of states, the screening wave vector can be written now as
$q_T=2 \pi e^2 n_h/\varepsilon kT$, where $\epsilon$ is the dielectric constant of the material.

Figure 4 illustrates the relationship between the  relaxation rates $1/\tau_{dh}^{*}$ and  $1/\tau_{hd}^{*}$  and temperature, represented by blue and magenta lines. The computations are carried out utilizing equations \ref{eq32}  with the parameter B set to 0.6.  Clearly, across a broad temperature range, the interaction rate maintains the relationship $1/\tau_{dh}^{*} > 1/\tau_{hd}^{*}$ for both regimes. Furthermore, the relaxation rate $1/\tau_{dh}^{*}$ surpasses $1/\tau_{dh}$ at temperatures exceeding 40K.

\section{Comparison with the experiment}
In sections V and VI we provide a theoretical description
of a 2D system with two types of carriers, one with a linear and another with
a quadratic spectrum. The rate of
the momentum transfer between these two types of carriers is found. We
consider a fully Fermi degenerate regime (FD) for both systems  at low
temperatures $(kT < \mu-\Delta, \mu)$ and a
partially degenerate regime (PD),  when the Dirac particles remain
degenerate, while the heavy holes obey the Boltzman statistic
$(\mu-\Delta < kT < \mu)$.

In a fully degenerate  (FD) regime we obtain the conventional expression for the
particle-particle collisions, slightly modified due to the difference in
spectrum and is given by equations \ref{eq30}.
Note that $1/\tau_{h d} << 1/\tau_{d h}$, and conventional $T^2$ behavior for $1/\tau_{int}\sim 1/\tau_{d h}$, characteristic of a particle with parabolic dispersion, becomes apparent.

In a partially degenerate (PD) regime $( \mu-\Delta < kT < \mu)$, the particle-particle collision rate $1/\tau_{dh,hd}^{*}$ shows a distinct temperature dependence.
By adjusting  expression for scattering rate while considering the constraint that interactions between holes conserve the overall momentum density it is given by equations \ref{eq32}. Because $1/\tau_{int}^{*}=1/\tau_{dh}^{*}+1/\tau_{hd}^{*} \sim1/\tau_{dh}^{*}$, it is expected that the temperature-dependent behavior of the relaxation rate can be described as follows: $1/\tau_{int}^{*}\sim T^{3.5}$.
\begin{table}[ht]
\caption{\label{tab2} Fitting parameters in equations 1-5 for 3 samples.}
\begin{ruledtabular}
\begin{tabular}{lccccccccc}
&Sample&$\tau_{h}$ & $\tau_{d}$ & $n_{h}$  & $n_{d}$ & A & B  \\
& & $10^{-13} $ & $10^{-13} $ &  $10^{11}$  &  $10^{11}$      \\
\hline
& & $s$  &  $s$ &  $ cm^{-2}$ & $ cm^{-2}$ &       \\
\hline
&A& $1.7$  & $12.2$ &  $4.78$ & $0.21$ &  1.2& 0.5   \\
&B& $0.94$  & $8.5$ &  $4.53$ & $0.22$ &  1.8& 0.6\\
&C& $1.2$  & $10.7$ &  $5.52$ & $0.23$ &  1.4  &  0.6\\
&A& $2.3$  & $3$ &  $0.28$ & $0.17$ &   & 0.15 \\
&B& $1.8$  & $3.5$ &  $0.53$ & $0.16$ &  & 0.15 \\
&C& $2.7$  & $3.55$ &  $0.28$ & $0.16$ &    & 0.16 \\
\end{tabular}
\end{ruledtabular}
\end{table}
\begin{figure}[ht]
\includegraphics[width=10cm]{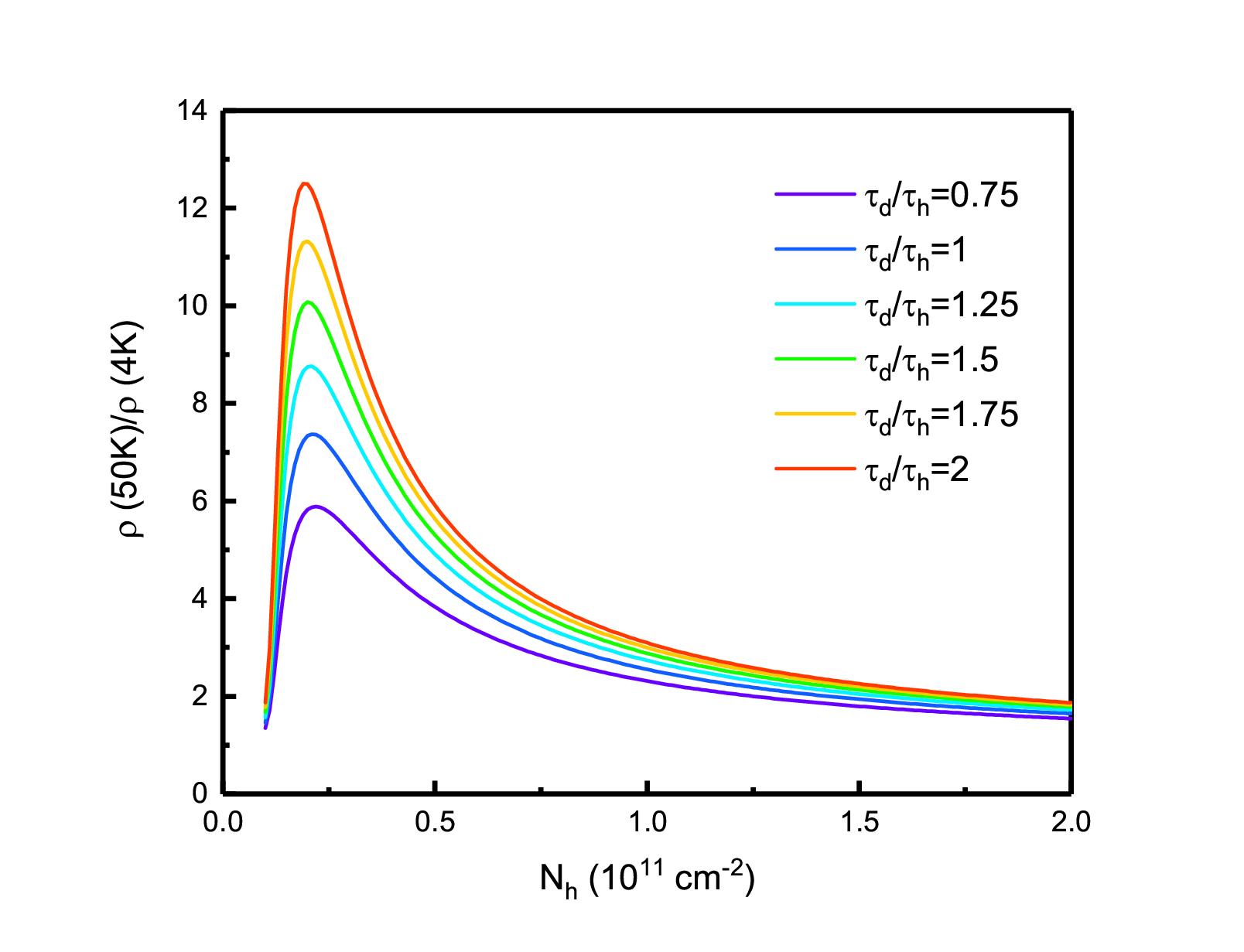}
\caption{(Color online)
 Ratio of resistivity at low an high temperature limits $\frac{\rho(T=50K)}{\rho(T=4.2K)}$ as a function of the density $N_{h}$  for different ratio $\tau_{d}/\tau_{h}$, calculated from eqs. \ref{eq9} and \ref{eq32}. Parameters are $\tau_{d}=3\times10^-13$ s, B=0.15. }
\end{figure}
In figure 3 we compare the results of the calculations with experimental
$\rho(T)$ dependencies using eqs. \ref{eq9} and \ref{eq30} for high and low total densities, corresponding to the different regimes, denoted as  FD and PD regimes withing corresponding models. We plot the experimentally measured
excess resistivity and that obtained theoretically $\Delta\rho(T)$ for three samples.
For comparison with the theory  we conducted a fitting analysis of the temperature-dependent data, as depicted in Figure 3, using a single adjustable parameters denoted as A and B. These parameters account for the strength of interaction between the Dirac and heavy holes, as described in equations \ref{eq9} and \ref{eq32}. The scattering parameters, denoted as $\tau_{h(d)}$, predominantly dictate the resistivity at lower temperatures. Importantly, adjusting these parameters within a reasonable range does not affect the friction coefficient, which is the key factor responsible for the temperature dependence of resistivity. In figs 3a,b,c  we  plot theoretical dependencies of the resistivity
excess for high total density ($N_{h}\approx 5\times 10^{11} cm^{-2}$).  Experimental data closely follows the expected dependence $\Delta\rho(T)\sim T^{2}$ for parameters  indicated in table II.

In figs 3d,e,f  we also plot theoretical dependencies of the resistivity
excess for low total density ($N_{h}\approx 0.5\times 10^{11} cm^{-2}$) using the PD model, which is in agreement with experimental data. Corresponding parameters are shown in the table I. It's important to emphasize that the expected behavior of $1/\tau_{dh}$ follows a power-law relationship with temperature $\sim T^{3.5}$, while resistivity exhibits a temperature-dependent relationship described by a power law of $\sim T^{\alpha}$ ($\alpha \approx 3$). This difference arises because resistivity demonstrates a saturation effect at low temperatures, attributed to scattering by impurities, and at high temperatures, it is influenced by various parameters, including the effective mass ratio as described in equation (1), leading to a temperature dependence resembling $\sim T^3$.

In Fig. 5, we illustrate the variation of the ratio $\rho(50K)/\rho(4.2K)$ with respect to the total density $N_h$ across various values of the scattering relaxation time ratio $\tau_d/\tau_h$, calculated from eqs.\ref{eq9} and \ref{eq32}. Notably, one observes that in realistic samples, this ratio converges towards the range of 6-12. This observation underscores a unique scenario in the realm of solid-state physics, wherein resistance is predominantly governed by interactions between particles. In stark contrast to the more typical scenarios where resistance is primarily influenced by disorder or phonon scattering, here, particle-particle collisions play a pivotal role, significantly exceeding the impact of impurity-related scattering.
At higher total densities and temperatures exceeding 30K, heavy holes enter into the Boltzmann regime, leading to an expected alteration in resistivity with changing temperature regimes. To verify this, we conducted a comparison between experimental data and calculations using the PD model. The outcomes of this comparison are illustrated in Figures 3a, 3b, and 3c. Notably, there exists a reasonable concordance with the PD model. However, owing to the limited temperature range, it proves challenging to experimentally discern between these two regimes at such densities. The variation in parameter B between high and low densities can be attributed to a deficiency in the model of short-range potential when it approaches the limit of the actual Coulomb potential.

One can see that the sample B seems to deviate from theoretical model at low temperatures (figure 3e). We attribute this deviation to density inhomogeneity, which may render our model less applicable at lower temperatures. However, it is noteworthy that at elevated temperatures, the dependence aligns with a $T^3$ relationship across a range exceeding one order of magnitude. Notably, we observe a $T^3$ dependence in samples A and C that adheres to this trend across more than three orders of magnitude, an occurrence not commonly observed in experiments.

\section {conclusion}
In conclusion, our investigation focused on the temperature-dependent resistivity in a gapless HgTe quantum well. We observed quadratic temperature dependencies arising from interactions between the Dirac and heavy holes in the fully degenerate regime. Conversely, a cubic temperature dependence emerged when heavy holes conformed to Boltzmann statistics while the Dirac holes retained Fermi liquid characteristics.

To validate our findings, we compared theoretical predictions with experimental data, revealing a satisfactory agreement. An interesting observation is that within the temperature range of 10-100 K, hole-hole scattering proves to be significantly more influential than impurity scattering. This is an uncommon occurrence, as in conventional metallic systems, particle-particle collisions typically do not limit conductivity.
\section{ACKNOWLEDGMENTS}
This work is supported  by FAPESP  (São Paulo Research Foundation )  Grants No. 2019/16736-2  and No 2021/12470-8,   CNPq (National Council for Scientific and Technological Development ),   and by   Ministry of Science and Higher Education of the Russian Federation and Foundation for the Advancement of Theoretical Physics and Mathematics "BASIS".

\end{document}